\documentstyle[a4,twocolumn]{article}

\newfont{\BB}{msbm10} \def\R{\mbox{\BB R}} 
\def\D{\mbox{\BB D}}

\makeatletter
\def\section{\@startsection {section}{1}{\z@}{-3.5ex plus -1ex minus
-.2ex}{2.3ex plus .2ex}{\large\bf}}
\def\subsection{\@startsection {subsection}{2}{\z@}{-3.25ex plus -1ex minus
-.2ex}{1.5ex plus .2ex}{\bf}}

\newcommand{\beq}{\begin{equation}}
\newcommand{\eeq}{\end{equation}}
\newcommand{\bea}{\begin{eqnarray}}
\newcommand{\eea}{\end{eqnarray}}
\newcommand{\nn}{\nonumber}

\newcommand{\proof}{\noindent {\bf Proof: }}
\def\qed{\hfill \vrule height 7pt width 7pt depth 0pt
          \medskip \\}

\newcommand{\bd}{\begin{displaymath}}
\newcommand{\ed}{\end{displaymath}}

\newfont{\bbb}{msbm8 scaled\magstep1} 

\def\bmat{\left[ \begin{array}}
\def\emat{\end{array} \right]}

\newcounter{acount}

\newtheorem{theo}{Theorem}

\newtheorem{remark}{Remark}

\begin{document}
\title{Steering a quantum system over a Schr\"{o}dinger bridge}

\author{Alessandro Beghi$^{\dag}$, Augusto Ferrante$^{\ddag}$, and Michele Pavon$^{\dag *}$ \smallskip
\\
 $^{\dag}$Dipartimento di  Elettronica e
Informatica, \\ Universit\`a di  Padova,\\ via Gradenigo 6/A, \\35131 Padova, Italy. \smallskip \\
$^{\ddag}$ Dipartimento di
Elettronica e Informazione,\\ Politecnico di
Milano,\\
 piazza Leonardo da Vinci 32,\\
20133 Milano, Italy. \smallskip\\
$^*$ LADSEB, CNR\\ C.so Stati Uniti, 4\\35100 Padova, Italy. \\
{\tt \{beghi,pavon\}@dei.unipd.it, ferrante@elet.polimi.it}}

\date{\ }

\maketitle

\thispagestyle{empty}
\begin{abstract}
We outline a new approach to the steering problem for quantum systems
relying on {\em Nelson's stochastic mechanics} and on the
{\em theory of Schr\"{o}dinger bridges}. The method is illustrated by
working out
a simple Gaussian example.
\end{abstract}

\section{Introduction}
The problem of steering a quantum system from a given state $\psi_0$ at time
$t_0$  to another given state $\psi_1$ at time $t_1$ applying an
external potential function lies at the heart of the field  of control for
quantum systems.
The applications are manyfold, including  control of molecular dynamics,
quantum computing,
nuclear magnetic resonance, quantum communications, etc., see
\cite{HT}-\cite{DD}, and references therein. In its entirety, the problem
consists in finding
a control potential function  $\{V_c(x,t);t_0\le t\le t_1\}$ in a suitable
class such that,
if $V_i(x,t)$ is the {\em ambient}  potential function, the solution
$\psi(x,t)$ of the
controlled Schr\"{o}dinger equation
\begin{eqnarray}\label{S} &&\frac{\partial{\psi}}{\partial{t}} =
\frac{i\hbar}{2m}\Delta\psi -
\frac{i}{\hbar}[V_i(x)+V_c(x,t)]\psi,\\ &&\psi(x,t_0)=\psi_0(x),
\end{eqnarray}
satisfies $\psi(x,t_1)=\psi_1(x)$. Of course, the first natural question to
pose  is the controllability question for the Schr\"{o}dinger equation.
Since this is a
distributed-parameter bilinear system, the problem is highly nontrivial,
and only very
limited results are so far available \cite{HT}. The problem can be
effectively studied for
N-level systems that approximate in a suitable way the infinite-dimensional
system \cite{RSDRP}.
\section{Elements of stochastic mechanics and Schr\"{o}dinger bridges}
In this paper, we outline a completely new approach to the steering problem
for the
Schr\"{o}dinger equation relying on {\em Nelson's stochastic mechanics}
\cite{N1}-\cite{P3} and on the {\em theory of Schr\"{o}dinger bridges}, cf.
\cite{S,F2} and
references therein. We remark, in passing, that this method has nothing to do
with the results of \cite{Z}-\cite{LK}, where the theory of Schr\"{o}dinger
bridges and
reciprocal processes was
employed to construct different versions of stochastic mechanics.

Nelson's stochastic mechanics is a quantization procedure for classical
dynamical systems based on diffusion processes. Given a quantum
evolution
$\{\psi(x,t);t_0\le t\le t_1\}$, where $x\in\R^n$, a $n$-dimensional Markov
diffusion process $\{x(t);t_0\le t\le t_1\}$,
called {\em Nelson's process}, is associated to it as follows. The
(forward) Ito differential
of $x(\cdot)$ is given by
\begin{eqnarray}\nonumber dx(t)&=&\left[\frac{\hbar}{m}\nabla \left(\Re
\log\psi(x(t),t)
+\Im \log\psi(x(t),t)\right)\right]dt\\
\label{N}
&+&\sqrt{\frac{\hbar}{m}}dw(t),
\end{eqnarray}
where $w$ is a standard, $n$-dimensional Wiener process. Moreover, the
probability density $\rho(\cdot,t)$ of
$x(t)$ satisfies
\begin{equation}\label{D}\rho(x,t)=|\psi(x,t)|^2,\quad \forall t \in [t_0,t_1].
\end{equation}
The construction of the Nelson
process corresponding to $\psi(x,t)$ in the case where
$\psi(x,t)$ vanishes requires considerable care. It will always be assumed
that $\psi$ is of
class $C^{2,1}$ with
\begin{equation}\label{FA}||\nabla\psi||^2_2\in L^1_{{\rm loc}}[[t_0,+\infty)].
\end{equation} This is Carlen's
{\em finite action condition}. Under these hypotheses, the Nelson measure
may be constructed,
\cite{C}, \cite [Chapter IV]{BCZ}, and references therein. Let us now
outline the theory of
the Schr\"{o}dinger bridges. The Kullback-Leibler pseudo-distance between
two probability
densities $p(\cdot)$ and $q(\cdot)$ is defined by

$$H(p,q):=\int_{\R^n}\log\frac{p(x)}{q(x)}p(x)dx.$$
This concept can be considerably generalized. Let $\Omega:={\cal
C}([t_0,t_1],\R^n)$, let
$W_x$ denote Wiener measure on $\Omega$ starting at $x$, and let
$$W:=\int W_x\,dx
$$
be stationary Wiener measure. Let $\D$ be the family of
distributions on $\Omega$ that are equivalent to $W$. For
$Q,P\in\D$, we define  the {\it relative
entropy } $H(Q,P)$ of $Q$ with respect to $P$ as
$$H(Q,P)=E_Q[\log\frac{dQ}{dP}].$$
It then follows from Girsanov's theorem that
\cite{F2}
\begin{eqnarray}\nonumber
&&H(Q,P)=H(q(t_0),p(t_0))+\\&&E_Q\left[\int_{t_0}^{t_1}\frac{1}{2}
[\beta^Q(t)-\beta^P(t)]\cdot
[\beta^Q(t)-\beta^P(t)]dt\right]\nonumber\\&&=H(q(t_1),p(t_1))+\nonumber\\&&E_Q
\left[\int_{t_0}^{t_1}\frac{1}{2}
[\gamma^Q(t)-\gamma^P(t)]\cdot [\gamma^Q(t)-\gamma^P(t)]dt\right].\nonumber
\end{eqnarray}
Here $q(t_0)$ is the marginal density of $Q$ at $t_0$, $\beta^Q$ and
$\gamma^Q$ are the
forward and the backward drifts of $Q$, respectively. Now let $\rho_0$ and
$\rho_1$ be two
everywhere positive probability densities. Let $\D(\rho_0,\rho_1)$
denote the set of distributions in $\D$ having the prescribed marginal
densities at $t_0$
and $t_1$. Given $P\in\D$, we consider the following problem:

$${\rm Minimize}\quad H(Q,P) \quad {\rm over} \quad \D(\rho_0,\rho_1).$$
This problem is connected through Sanov's theorem to a problem of large
deviations  of the empirical distribution, according to Schr\"{o}dinger
original motivation
\cite{S,F2}. If there is at least one $Q$ in $\D(\rho_0,\rho_1)$ such that
$H(Q,P)<\infty$,
it may be shown that there exists a unique minimizer $Q^*$ in
$\D(\rho_0,\rho_1)$ called
{\em the Schr\"{o}dinger bridge} from $\rho_0$ to $\rho_1$ over $P$. If
(the coordinate
process under) P is Markovian with forward drift field $b_+^P(x,t)$ and
transition density
$p(\sigma,x,\tau,y)$, then $Q^*$ is also Markovian with forward drift field
$$b_+^{Q^*}(x,t)=b_+^P(x,t)+\nabla\log\phi(x,t),$$ where the function
$\phi$ solves together
with another function $\hat{\phi}$ the system \begin{eqnarray}\nonumber
&&\phi(t,x)=\int p(t,x,t_1,y)\phi(t_1,y)dy,\\\nonumber&&\hat{\phi}(t,x)=\int
p(t_0,y,t,x)\hat{\phi}(t_0,y)dy
\end{eqnarray}
with the boundary conditions
$$\phi(x,t_0)\hat{\phi}(x,t_0)=\rho_0(x),\quad
\phi(x,t_1)\hat{\phi}(x,t_1)=\rho_1(x). $$
Moreover, we have  $\rho(x,t)=\phi(x,t)\hat{\phi}(x,t), \forall t\in
[t_0,t_1]$.
\section{Steering for quantum systems}
We now
show that the theory of Schr\"{o}dinger bridges can be employed, jointly
with stochastic mechanics,
to attack the
steering problem for quantum systems. First of all, observe
that everything we said about Schr\"{o}dinger bridges continues to hold if
we consider
finite-energy diffusions with diffusion coefficient equal to
$\frac{\hbar}{m}$ rather than
$1$. Let $\psi_0$ and $\psi_1$ be the given initial and
final quantum states. Consider a  {\em reference} quantum evolution
$\{\psi(x,t);t_0\le t\le t_1\}$ solving the  Schr\"{o}dinger equation
$$\frac{\partial{\psi}}{\partial{t}} = \frac{i\hbar}{2m}\Delta\psi -
\frac{i}{\hbar}V(x)\psi,
$$
and satisfying Carlen's finite action condition (\ref{FA}). Let $P\in\D$ be
the Markovian measure
of the Nelson process associated to $\{\psi(x,t)\}$ as in (\ref{N})-(\ref{D}).
Hence, in particular,
the probability density satisfies $\rho(x,t)=|\psi(x,t)|^2$. Thus, if we
write
$$\psi(x,t)=\exp [R(x,t)+\frac{i}{\hbar}S(x,t)],$$
the forward drift of the Nelson process
is then given by
$$b_+^P(x,t)=\frac{1}{m}\nabla S(x,t)+\frac{\hbar}{m}\nabla R(x,t).$$
\noindent
We then have the following result.
\begin{theo}
Let $Q^*$ be the {\em  Schr\"{o}dinger bridge} from $|\psi_0|^2$ to
$|\psi_1|^2$ over $P$
(see previous section). Then, $Q^*$ has forward drift field
$$b_+^{Q^*}(x,t)=\frac{1}{m}\nabla S(x,t)+\frac{\hbar}{m}\nabla
R(x,t)+\frac{\hbar}{m}\nabla\log\phi(x,t),$$ where the function $\phi$
solves together
with another function $\hat{\phi}$ the system
\begin{eqnarray}\label{p1}
&&\frac{\partial{\phi}}{\partial{t}}+(\frac{1}{m}\nabla S +
\frac{\hbar}{m}\nabla R)\cdot\nabla\phi + \frac{\hbar}{2m}\Delta\phi=0,\\
\label{p2}&&\frac{\partial{\hat{\phi}}}{\partial{t}}+\nabla\cdot
\left[(\frac{1}{m}\nabla S +
\frac{\hbar}{m}\nabla R)\hat{\phi}\right] -
\frac{\hbar}{2m}\Delta\hat{\phi}=0,
\end{eqnarray}
with the boundary conditions
$$\phi(x,t_0)\hat{\phi}(x,t_0)=|\psi_0|^2(x),\quad
\phi(x,t_1)\hat{\phi}(x,t_1)=|\psi_1|^2(x). $$
The one-time probability density of  $Q^*$ satisfies
$$\tilde{\rho}(x,t)=\phi(x,t)\hat{\phi}(x,t).$$
Define, for $t\in [t_0,t_1]$,
\begin{eqnarray}&&\label{F}\tilde{S}(x,t)=S(x,t)+\hbar
R(x,t)+\frac{\hbar}{2}\log
\frac{\phi(x,t)}{\hat{\phi}(x,t)},\\\label{R}
&&\tilde{R}(x,t)=\frac{1}{2}\log \tilde{\rho}(x,t).
\end{eqnarray}
Let  $\{\tilde{\psi}(x,t);t_0\le t\le t_1\}$ be defined by
$$ \tilde{\psi}(x,t)=\exp [\tilde{R}(x,t)+\frac{i}{\hbar}\tilde{S}(x,t)].
$$
Then,  $\{\tilde{\psi}(x,t)\}$ solves the Schr\"{o}dinger equation
(\ref{S}) with
controlling potential function $V_c(x,t)$ given by
\begin{eqnarray}\nonumber
V_c(x,t)&=&V(x)-V_i(x)+\\&&\frac{\hbar^2}{m}\left[\frac{\Delta
\sqrt{\tilde{\rho}(x,t)}}
{\sqrt{\tilde{\rho}(x,t)}}-\frac{\Delta \sqrt{\rho(x,t)}}
{\sqrt{\rho(x,t)}}\right],\label{V}
\end{eqnarray}
and we have
$$|\tilde{\psi}(x,t_0)|=|\psi_0(x)|,\quad|\tilde{\psi}(x,t_1)|=|\psi_1(x)|.
$$
Moreover, the {\em the Schr\"{o}dinger bridge} $Q^*$ is indeed the Nelson
process
associated to the new quantum evolution  $\{\tilde{\psi}(x,t);t_0\le t\le
t_1\}$.
\end{theo}
\proof
As is well known \cite{N1}, $R$ and $S$ satisfy the system of nonlinear
p.d.e.'s
\begin{eqnarray}\label{sa1}
\frac{\partial R}{\partial{t}}+\frac{1}{m}\nabla R\cdot\nabla S+
\frac{1}{2m}\Delta S=0,\\\label{sa2}
\frac{\partial S}{\partial t}+\frac{1}{2m}\nabla S\cdot\nabla S+ V
-\frac{\hbar}{2m}\left[\nabla R\cdot\nabla R+\Delta R\right]=0.
\end{eqnarray}
A long calculation employing definitions (\ref{F})-(\ref{R}), and equations
(\ref{p1}), (\ref{p2}), (\ref{sa1}) and (\ref{sa2}) establishes (\ref{V}).
The last assertion
follows at once observing that (\ref{F})-(\ref{R}) imply
$$b_+^{Q^*}(x,t)=\frac{1}{m}\nabla \tilde{S}(x,t)+\frac{\hbar}{m}\nabla
\tilde{R}(x,t).
$$
\qed

\begin{remark} {\em Notice that when $V(x)=V_i(x)$, namely the reference
quantum evolution
takes place in the ambient potential $V_i$, the controlling potential
reduces to
$$V_c(x,t)=\frac{\hbar^2}{m}\left[\frac{\Delta \sqrt{\tilde{\rho}(x,t)}}
{\sqrt{\tilde{\rho}(x,t)}}-\frac{\Delta \sqrt{\rho(x,t)}}
{\sqrt{\rho(x,t)}}\right].
$$
Instead, when the ambient potential is zero, we get the following
remarkable invariance property:
$$V_c(x,t)-\frac{\hbar^2}{m}\frac{\Delta \sqrt{\tilde{\rho}(x,t)}}
{\sqrt{\tilde{\rho}(x,t)}}=V(x)-\frac{\hbar^2}{m}\frac{\Delta \sqrt{\rho(x,t)}}
{\sqrt{\rho(x,t)}}.
$$
The quantity
$$-\frac{\hbar^2}{2m}\frac{\Delta
\sqrt{\tilde{\rho}(x,t)}}{\sqrt{\tilde{\rho}(x,t)}}
$$
is called {\em quantum potential} in quantum mechanics \cite{G,BH}because
it appears in
the Hamilton-Jacobi like equation (\ref{sa2}).}
\end{remark}
\begin{remark}
{\em The quantum evolution  $\{\tilde{\psi}(x,t);t_0\le
t\le t_1\}$ has, by construction, the desired absolute values at times
$t_0$ and $t_1$.
Notice that any choice of the ``reference process" $P$ produces a quantum
evolution with these properties, and a corresponding control potential
function. Thus, we
can try to choose $P$ (our ``free parameter"!) so that the phase function
$\tilde{S}(x,t)$ has the
prescribed initial and final values, thereby achieving the desired
steering, see the
example in the following section.}
\end{remark}

\section{Example}
The aim is that of shifting the mean of a Gaussian wave packet, that
is, passing from the initial quantum
state
\bea
\psi_0(x)&=&\left (\frac{\omega}{\pi}\right)^{1/4} \exp(-\frac{\omega
x^2}{2})\nn\\
&=&\exp(R_0(x)+iS_0(x))
\eea
at $t=0$ to the final quantum state
\bea
\psi_1(x)&=&\left (\frac{\omega}{\pi}\right)^{1/4} \exp(-\frac{\omega
(x-1)^2}{2}) \nn\\
&=&\exp(R_1(x)+iS_1(x))
\eea
at $t=1$. As we shall show elsewhere \cite{BFP}, to solve this simple
problem a
direct approach is actually feasible. We choose here to solve the problem
via the theory of Schr\"{o}dinger bridges, in order to illustrate
the method of the previous section. For notational convenience, let us
assume that we are in  a
reference system where $m=\hbar=1$
and that $\omega=\pi$. We will  often omit in the sequel the function
arguments $t$ and $x$. Let us take as
reference evolution
\beq
\psi(x,t)=\exp(R(x,t)+iS(x,t))\:,\label{refev}
\eeq
where
\bea
R(x,t)&=&-\frac{\omega}{2}(x-m(t))^2\\
S(x,t)&=&cx+d(t)\:,
\eea
with
\bea
m(t)&=&m_1+m_2t\\
c&=&\dot m(t)=m_2
\eea
and $d$ is to be specified. The Schr\"odinger system (\ref{p1})-(\ref{p2}) is given by the
following two equations for $\phi$ and $\hat \phi$
\bea
\frac{\partial \phi}{\partial t} + \left ( \frac{\partial
S}{\partial x} + \frac{\partial R}{\partial x}
\right ) \frac{\partial \phi}{\partial
x}+\frac{1}{2}\frac{\partial^2 \phi}{\partial x^2} &=& 0
\label{s1}
\\
\frac{\partial \hat \phi}{\partial t} + \frac{\partial}{\partial
t}\left [ \left ( \frac{\partial
S}{\partial x} +
\frac{\partial R}{\partial x}
\right ) \hat \phi \right ]-\frac{1}{2}\frac{\partial^2
\hat \phi}{\partial x^2} &=& 0 \label{s2}
\eea
with boundary conditions
\bea
\phi(x,0)\hat \phi(x,0)&=&\rho_0(x)=\exp(-\omega x^2) \label{bc1}\\
\phi(x,1)\hat \phi(x,1)&=&\rho_1(x)=\exp(-\omega (x-1)^2)\label{bc2}
\eea
Assuming that
\bea
\phi(x,t)&=&\exp(\alpha(t)x^2+\beta(t)x+\gamma(t)) \label{fi}\\
\hat \phi(x,t)&=&\exp(\hat \alpha(t)x^2+\hat \beta(t)x+\hat
\gamma(t))\label{fihat}\:,
\eea
we find the following sets of equations for $\alpha, \hat \alpha$,
$\beta, \hat \beta$, and $\gamma,
\hat \gamma$:
\beq
\left \{ \begin{array}{lcl}
\dot \alpha-2\alpha \omega+2\alpha^2&=&0\\
\dot \beta+2\alpha c+2 \alpha \omega m - \omega \beta +2\alpha \beta &=& 0\\
\dot \gamma + c \beta +\omega \beta m +\frac{1}{2}\beta^2+\alpha &=&0
\end{array}\right.\label{sis1}
\eeq
and
\beq
\left \{ \begin{array}{lcl}
\dot {\hat \alpha}-2\hat \alpha \omega-2\hat \alpha^2&=&0\\
\dot {\hat \beta}+2\hat \alpha c+2 \hat \alpha \omega m - \omega \hat
\beta -2\hat \alpha \hat \beta &=&
0\\
\dot{\hat  \gamma}-\omega + c \hat \beta +\omega \hat \beta m
-\frac{1}{2}\hat \beta^2-\hat \alpha &=&0
\end{array}\right.\:.\label{sis2}
\eeq
The boundary conditions  (\ref{bc1})-(\ref{bc2}) yield the
constraints on the values of $\alpha, \hat \alpha$, $\beta, \hat
\beta$, and $\gamma,
\hat \gamma$ at $t=0,1$:
\beq
\left \{
\begin{array}{lcl}
\alpha_0+\hat \alpha_0 &=&-\omega\\
\beta_0+\hat \beta_0&=&0\\
\gamma_0+\hat \gamma_0&=&0
\end{array}
\right. \label{sis3}
\eeq
and
\beq
\left \{
\begin{array}{lcl}
\alpha_1+\hat \alpha_1 &=&-\omega\\
\beta_1+\hat \beta_1&=&2\omega\\
\gamma_1+\hat \gamma_1&=&-\omega
\end{array}
\right.\:.\label{sis4}
\eeq
It is easy to see that 
\bea
\alpha(t)&\equiv& 0 \label{a1}\\ 
\hat \alpha (t) &\equiv&
-\omega\label{a2}\\ 
\beta (t) &=& \beta_0 e^{\omega t}\label{a3}
\eea satisfy
(\ref{sis1})-(\ref{sis4}). The value of $\hat
\beta (t)$ can be found by integration
\beq
\hat \beta (t) = 2\omega m(t)-e^{-\omega t}(\beta_0+2\omega m_1)\:. \label{betahat}
\eeq
Using the relation $\beta_1+\hat \beta_1=2\omega$, we obtain
\beq
\beta_0=2\omega\frac{1+m_1e^{-\omega}-(m_1+m_2)}{e^\omega-e^{-\omega}}
\:.\label{beta0}
\eeq
Given $\beta$ and $\hat \beta$, it is possible to obtain $\gamma$ and
$\hat \gamma$ (by integration),
where $\gamma_0$ is so far to be specified. Thus, we get
\bea
\gamma(t)&=&\gamma_0+\frac{\beta_0^2}{4\omega}(1-e^{2\omega
t})+\beta_0(m_1-e^{\omega t}m(t)) \label{gamma}\\
\hat \gamma(t)&=&-\gamma_0+\frac{1}{4\omega} [ \beta_0^2(1-e^{-2\omega t})\nn\\
&&+4 \beta_0\omega e^{-2\omega t}(e^{\omega t}m(t)-m_1)\nn\\
&&+4\omega^2e^{-2\omega t}(e^{\omega t}m(t)-m_1)^2]\label{gammahat}
\eea
It can be checked that the constraint
$\gamma_1+\hat
\gamma_1=-\omega$ is always satisfied, whatever the value of
$\gamma_0$. Matching of the phase at $t=0,1$
requires that
\bea
S(x,0)+R(x,0)+\ln\:\phi(x,0)&=&S_0(x)+\frac{1}{2}\ln\:\rho_0(x)\nn\\
S(x,1)+R(x,1)+\ln\:\phi(x,1)&=&S_1(x)+\frac{1}{2}\ln\:\rho_1(x)\nn\:,
\eea
or equivalently that
\bea
m_2+\omega m_1+\beta_0 &=& 0 \label{u1}\\
\gamma_0-\frac{\omega}{2}m_1^2 +d_0&=&0\label{u2}\\
m_2+\omega(m_1+m_2)+\beta_1-\omega &=&0\label{u3}\\
\frac{\omega}{2}(m_1+m_2)^2+\gamma_1+\frac{\omega}{2}+d_1&=&0\:,\label{u4}
\eea
where $d_0$ and $d_1$ are the values of $d(t)$ at $t=0,1$. Some
computations show that eqns. (\ref{beta0})
and (\ref{u1})-(\ref{u4}) are satisfied by the following values
\bea
m_1&=&-\frac{e^\omega-1}{2-2e^\omega+\omega+\omega e^\omega}\label{41}\\
m_2&=&\frac{\omega(e^\omega+1)}{2-2e^\omega+\omega+\omega e^\omega}\\
\beta_0&=&-\frac{2\omega}{2-2e^\omega+\omega+\omega e^\omega}\\
\gamma_0&=&\frac{\omega}{2}\frac{(e^\omega-1)^2}{(2-2e^\omega+\omega+
\omega e^\omega)^2}\\
d_0&=&0\\
d_1&=&-\frac{\omega(1+e^\omega)(1-e^\omega+\omega+\omega
e^\omega)}{(2-2e^\omega+\omega+\omega
e^\omega)^2}\:.
\eea
A possible choice for $d(t)$ is then
\beq
d(t)=d_1t\:. \label{47}
\eeq
By choosing $m$, $d$, $\alpha, \hat \alpha$, $\beta, \hat
\beta$, and $\gamma,
\hat \gamma$ according to equations (\ref{a1})-(\ref{betahat}),  (\ref{gamma})-(\ref{gammahat}), (\ref{41})-(\ref{47}), both the
reference evolution (\ref{refev}) and the solution $(\phi, \hat \phi)$ of the Schr{\"o}dinger system (\ref{s1})-(\ref{bc2}) are 
completely specified. The expressions of the controlled quantum evolution $\tilde \psi(x,t)$ and of the controlling potential function
$V_c(x,t)$ can then be derived as in Theorem 1.

\end{document}